\documentclass[aps,prb,superscriptaddress,twocolumn,reprint,10pt,floatfix]{revtex4-1}
\usepackage{amsfonts}
\usepackage{amstext}
\usepackage{amssymb}
\usepackage{amsmath}
\usepackage[utf8]{inputenc}
\usepackage{graphicx}
\usepackage{color,xcolor}

\usepackage{hyperref}
\hypersetup{colorlinks=true,citecolor=blue,linkcolor=blue,filecolor=blue,urlcolor=blue,pdfstartview=FitH,pdfpagemode=UseNone}

\bibpunct{[}{]}{,}{n}{}{}

\begin{document}

\title{An efficient method for computing the electronic transport properties of a multi-terminal system}

\author{Leandro R. F. Lima}
\affiliation{Instituto de F\'isica, Universidade Federal Fluminense, 24210-346 Niter\'oi, RJ, Brazil}

\author{Amintor Dusko}
\affiliation{Instituto de F\'isica, Universidade Federal Fluminense, 24210-346 Niter\'oi, RJ, Brazil}

\author{Caio Lewenkopf}
\affiliation{Instituto de F\'isica, Universidade Federal Fluminense, 24210-346 Niter\'oi, RJ, Brazil}

\date{\today}

\begin{abstract}
We present a multiprobe recursive Green's function method to compute the transport properties of mesoscopic systems
using the Landauer-B\"uttiker approach.
By introducing an adaptive partition scheme, we map the multiprobe problem into the standard two-probe recursive Green's
function method.
We apply the method to compute the longitudinal and Hall resistances of a disordered graphene sample, a system of current
interest. We show that the performance and accuracy of our method compares very well with other state-of-the-art schemes.
\end{abstract}

\maketitle

%%%%%%%%%%%%%%%%%%%%%%%%%%%%%%%%%%%%%%%%%%%%%%%%%%%%
\section{Introduction}
\label{sec:intro}
%%%%%%%%%%%%%%%%%%%%%%%%%%%%%%%%%%%%%%%%%%%%%%%%%%%%

The recursive Green functions (RGF) method is a powerful tool to calculate the electronic transport properties of
quantum coherent mesoscopic systems \cite{Thouless1981,MacKinnon1985,Sols1989,Baranger1991}.
Several important improvements have been proposed over the last decades to improve the method performance, 
like an optimal block-diagonalization scheme \cite{Wimmer2009} and a modular RGF method \cite{Rotter2000,Libisch2012}, 
to name a few. 
Notwithstanding, with few exceptions so far the method has been mainly used to compute the Landauer conductance 
in two-terminal devices, that is, in systems attached to two leads in contact with electronic reservoirs.

Some studies \cite{Baranger1988,Baranger1990,Baranger1991,Kazymyrenko2008,Yang2011,Thorgilsson2014} 
have extended the method to treat multi-probe systems. However, the latter are designed to address systems with 
very simple geometries, except for Ref.~\cite{Kazymyrenko2008} at the expense of increasing the algorithm complexity.

In this paper we report a multi-probe recursive Green's function (MPRGF) method that generalizes and improves the
previous developments.
Our scheme is simple to implement, very flexible and capable of addressing systems with arbitrary
geometry, and shows a superior or similar performance as compared to the others.

This paper is organized as follows: In Sec.~\ref{sec:multiprobe} we summarize the multi-probe Landauer-B\"uttiker
approach and present expressions for the observables of interest cast in terms of Green's functions.
In Sec.~\ref{sec:method} we introduce the adaptive partition scheme that allows for an efficient solution of the problem.
We illustrate the method using a simple pedagogical model.
Section~\ref{sec:application} shows an application of the MPRGF method in a physical system of current interest.
The processing time and accuracy of the method are discussed in Sec.~\ref{sec:benchmark}.
We summarize our results in Sec.~\ref{sec:discussion}.

%%%%%%%%%%%%%%%%%%%%%%%%%%%%%%%%%%%%%%%%%%%%%%%%%%%%
\section{Electronic transport properties in multiprobe systems}
\label{sec:multiprobe}
%%%%%%%%%%%%%%%%%%%%%%%%%%%%%%%%%%%%%%%%%%%%%%%%%%%%

In this Section we present in a nutshell the main results of the Landauer-B\"uttiker approach to calculate the
transport properties of a multi-probe quantum coherent mesoscopic system.
The RGF method can be implemented for both a finite element discretization of the Schr\"odinger equation
\cite{Datta1995,Ferry2009} or a tight-binding model based on a linear combination of atomic orbitals
\cite{Lewenkopf2013,Ridolfi2017}.
For simplicity, in this paper we consider nearest neighbor tight-binding models that use a single orbital per site.
The generalization to more realistic models is straightforward.
With this restriction, we can use the same discrete notation for both above mentioned Hamiltonian models.

%----------------------------------------------------------- F I G U R E 1
\begin{figure}[tbp]
	\centering
		\includegraphics[width=.85\columnwidth]{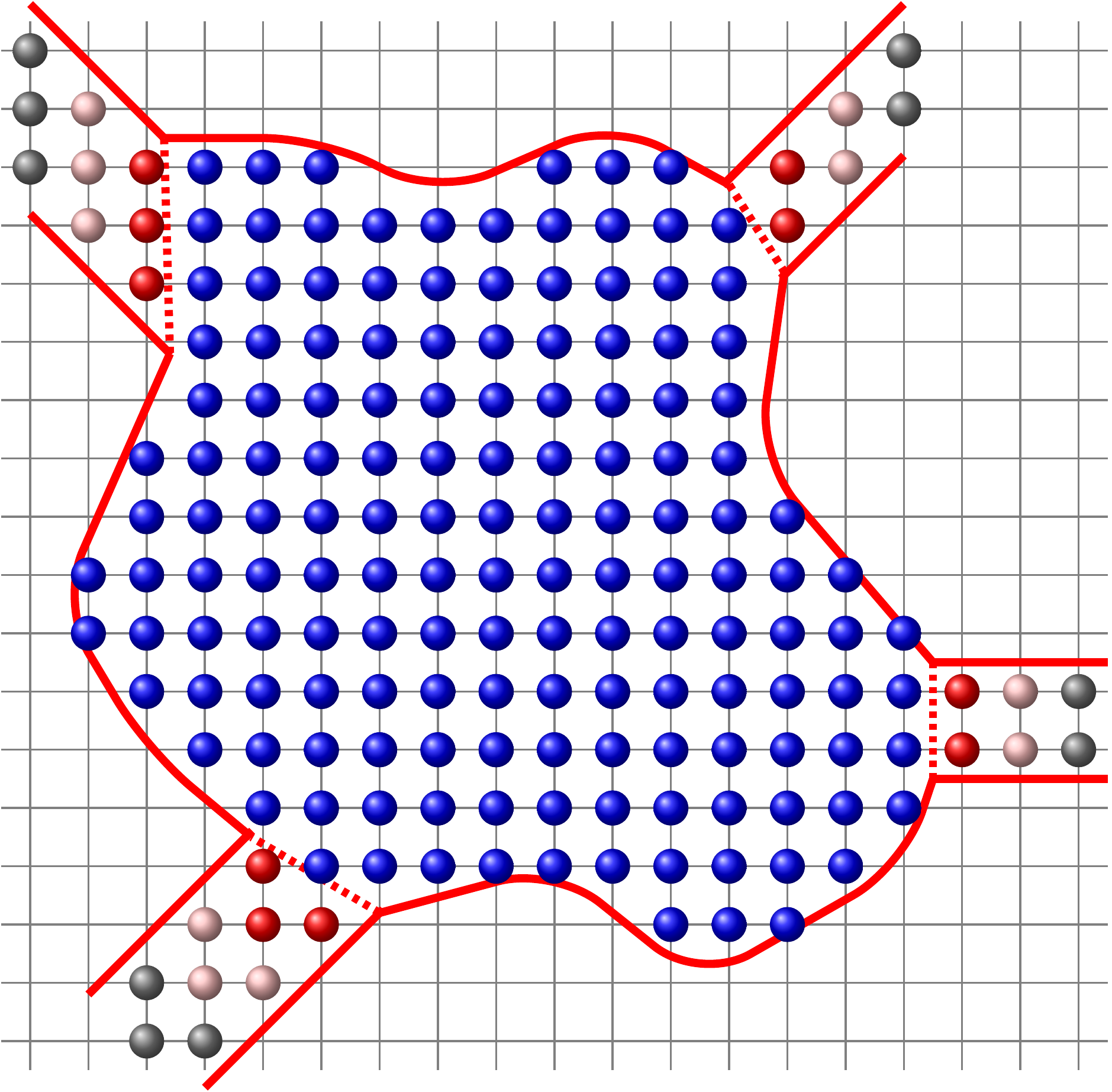}
	\caption{
Sketch of the mesoscopic system of interest.
Blue sites compose the central region.
The leads $\alpha=1,\cdots,N_L$ are connected to electronic reservoirs in thermal and chemical equilibrium that act as
terminals. The dashed lines indicate the lead--central region interfaces. Red, pink and gray sites represent the first, 
second and third primitive unit cells of different leads.
	}
	\label{fig:generalsystem}
\end{figure}
%----------------------------------------------------------- F I G U R E 1

In linear response theory, the multiterminal Landauer-B\"uttiker formula for the electronic current $I_\alpha$ at the terminal
$\alpha$ reads \cite{Buttiker1985,Buttiker1986,Ihn2010}, see Fig.~\ref{fig:generalsystem},
\begin{align}
I_\alpha = \sum_{\beta=1}^{N} \mathcal G_{\alpha\beta} \left(V_\alpha-V_\beta \right),
\label{current}
\end{align}
where the greek letters label the terminals, $V_\alpha$ is the voltage applied to the $\alpha$-terminal and $G_{\alpha\beta} $
is the conductance given by
\begin{align}
\mathcal G_{\alpha \beta} = \frac{2e^2}{h}\int_{-\infty}^{\infty} dE \left(-\frac{\partial f}{\partial E} \right) T_{\alpha\beta}(E)
\label{conductance}
\end{align}
that is cast in terms of the the Fermi distribution  $f(E) = [1 + e^{(E-\mu)/k_B T}]^{-1}$ and the transmission
$T_{\alpha\beta} (E)$. The factor $2$ assumes spin degeneracy. For cases where the system Hamiltonian
depends explicitly on the electron spin projection, one incorporates this degree of freedom in the lattice basis,
doubling its size.

The transmission $T_{\alpha\beta} (E)$ is given by \cite{MeirWingreen1992}
\begin{align}
\label{eq:transmission}
T_{\alpha\beta} (E)= {\rm tr} \left[ \mathbf \Gamma_\beta(E) \mathbf G^r(E) \mathbf \Gamma_\alpha(E) \mathbf G^a(E) \right]
\end{align}
where $\mathbf G^r=\left(\mathbf G^a\right)^\dagger$ is the retarded Green's function of the full system
(central region and leads, see Fig.~\ref{fig:generalsystem}), whose computation is the central goal of this paper,
while $\mathbf \Gamma_\beta$ is the linewidth of the lead corresponding to the $\beta$-terminal.
Both $\mathbf G^r$ and $\mathbf \Gamma_\beta$  are expressed in a discrete representation, while
$\mathbf G^r$ has the dimension of the number of sites in the central region, the dimension of
$\mathbf \Gamma_\beta$ is the number of sites at the $\beta$-lead-central region interface.
Following the standard prescription \cite{Ferry2009,Lewenkopf2013}, the leads are considered as semi-infinite.
The decay width is related to the embedding self-energy, namely
\begin{align}
\mathbf	\Sigma_\alpha = \mathbf V^\dagger \mathbf G^r_\alpha \mathbf V
\end{align}
and
\begin{align}
\mathbf	\Gamma_\alpha = -2\,\text{Im}\left(\mathbf \Sigma_\alpha\right),
\end{align}
where $\mathbf V$ gives the coupling matrix elements between the lead $\alpha$ and the central region,
and $\mathbf G^r_\alpha$ is a contact Green's function that casts the electron dynamics in the leads, which
can be calculated in a number of ways \cite{LopezSancho1985, MacKinnon1985, Rocha2006, Wimmer2009thesis}.

The local density of states (LDOS) can be directly obtained from $\mathbf G^r$, namely
\begin{align}
	\rho(j) = -\frac{1}{\pi}\, \text{Im}\,G^r_{jj},
	\label{eq:ldos}
\end{align}
where $j$ corresponds to the site at $\mathbf r_j$.

Another important quantity of interest is $T_{jj'}^\alpha$ the local transmission between two sites $j$ and $j'$
due to electrons injected from the terminal $\alpha$, namely \cite{Cresti2003,Lewenkopf2013}
\begin{align}
\label{eq:local_transmission}
T_{jj'}^\alpha = -2{\rm  Im }\left\{ \left[ \mathbf G^r \mathbf \Gamma_\alpha \mathbf G^a\right]_{j,j'}H_{j',j}\right\},
\end{align}
where $\mathbf H$ is the system Hamiltonian in the discrete representation.

%%%%%%%%%%%%%%%%%%%%%%%%%%%%%%%%%%%%%%%%%%%%%%%%%%%%
\section{Adaptive slicing scheme}
\label{sec:method}
%%%%%%%%%%%%%%%%%%%%%%%%%%%%%%%%%%%%%%%%%%%%%%%%%%%%

\begin{figure}[tbp]
	\centering
		\includegraphics[width=.8\columnwidth]{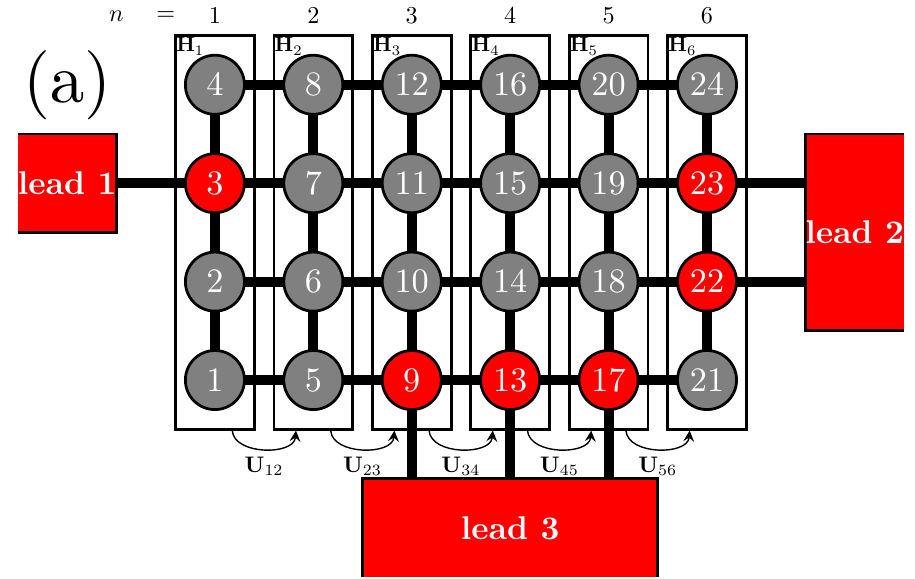}
		\includegraphics[width=.8\columnwidth]{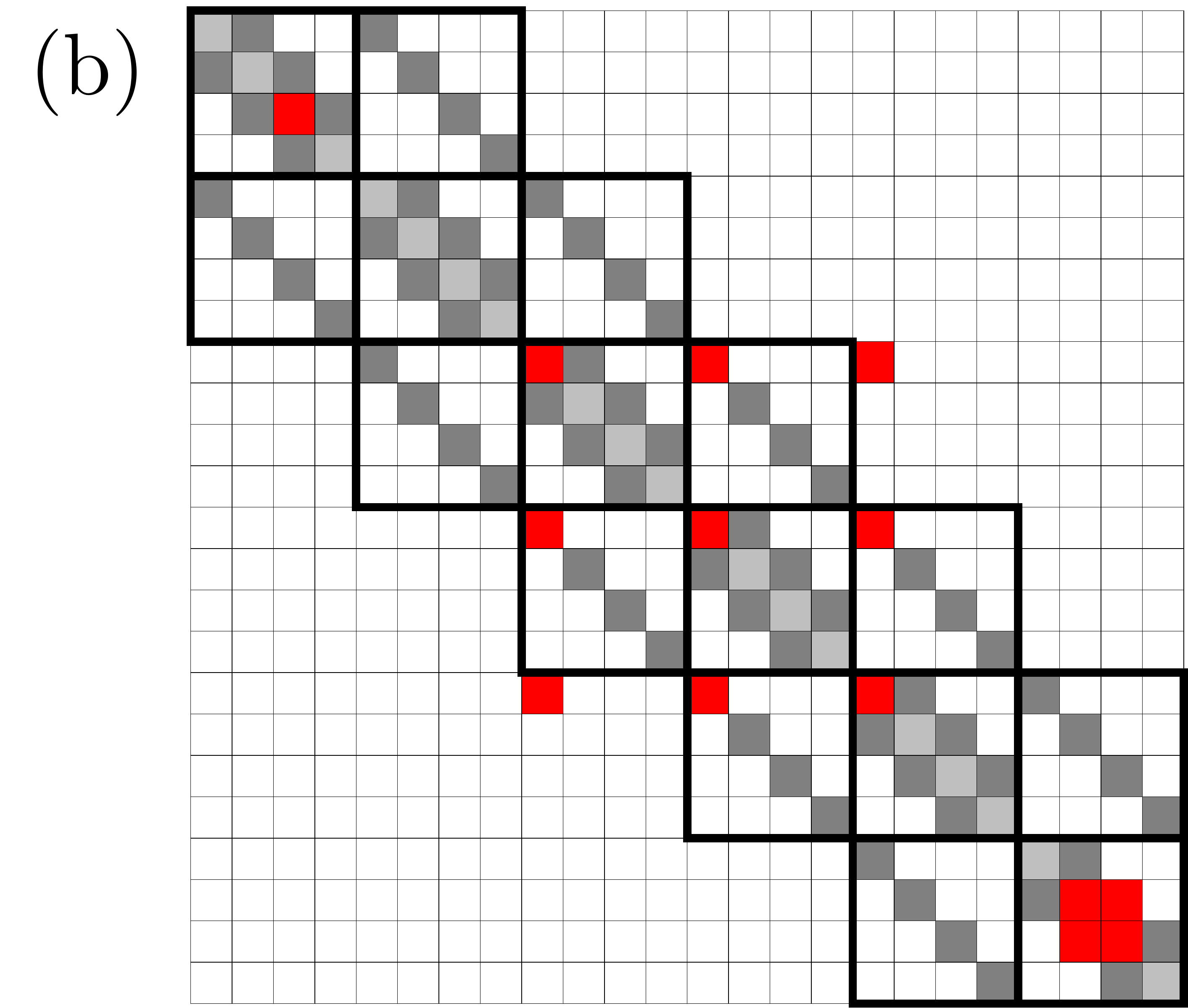}
	\caption{
(a) Standard RGF slicing scheme applied to a three-probe system with $N=6$ slices with intra-slice hopping matrices $\mathbf H_n$ and inter-slice hopping matrices $\mathbf U_{n,n+1}$ for $n=1,\cdots,6$.
Each slice $n$ has $M_n=4$ sites. % that determine the dimension of the hopping matrices.
(b) Matrix structure of $\mathbf H+\mathbf \Sigma$ for the system (a). White boxes correspond to $H_{ij} + \Sigma_{ij}=0$ and the 
red ones to matrix elements where $\Sigma_{ij}\neq0$.
The thick solid lines highlight the intra- and inter-slice blocks.
	}
	\label{fig:slicing_scheme_two_probe}
\end{figure}

In this section we put forward an efficient adaptive slicing scheme tailor-made for multi-terminal systems.
We present general expressions for the Green's functions and illustrate how the method works using a small and very
simple lattice model, depicted in Fig.~\ref{fig:slicing_scheme_two_probe}, which serves as a practical guide for the
system labels we use.
In what follows we deal only with retarded Green's functions $\mathbf G^r$, where $E\rightarrow E + i \eta$. Hence, to simplify 
the notation, from now on we omit the superindex $r$. 

The implementation of the RGF method requires a partition of the system into $N$ domains or ``slices''.
A given slice $n$, that contains $M_n$ sites, is connected only with the slice $n-1$ and the slice $n+1$ through
the hopping matrices $\mathbf U_{n,n-1}$ and $\mathbf U_{n,n+1}$, respectively, and has an internal hopping
matrix $\mathbf H_n$.
See the lattice model in Fig.~\ref{fig:slicing_scheme_two_probe} for details.
Several partition schemes have been proposed in the literature \cite{Wimmer2009,Yang2011,Thorgilsson2014}.
As a rule, it is preferable to minimize the number of sites $M_n$ inside each slice $n$ and increase the number
of slices $N$, since the computational time cost scales as $N\times M_n^3$.

For two-terminal geometries, it is convenient to connect the
first $n=1$ and the last $n=N$ slices to the left lead $L$ and to the right lead $R$, respectively.
This partition scheme leads to a block tridiagonal Hamiltonian and a retarded self-energy
$\mathbf \Sigma$ coupled only to the first and last slices of the system.
Thus, in a block matrix representation the self-energy has the form $\mathbf \Sigma_{n,n'} =
\mathbf \Sigma_{1,1}\delta_{n,1}\delta_{n',1} + \mathbf \Sigma_{N,N}\delta_{n,N}\delta_{n',N}$ and
$\mathbf{H} + \mathbf \Sigma$ has a block tridiagonal structure.
As long as the last requirement is met, one can apply the RGF method straightforwardly.

Unfortunately, this simple scheme does not work for setups with more than two terminals.
Figure~\ref{fig:slicing_scheme_two_probe} shows the standard RGF slicing scheme applied to 
a very simple two-dimensional lattice system with three-terminals.
We show the lattice model in Fig.~\ref{fig:slicing_scheme_two_probe}(a) and the corresponding matrix 
structure of $\mathbf H+\mathbf \Sigma$ for a nearest-neighbor coupling model Hamiltonian in Fig.~\ref{fig:slicing_scheme_two_probe}(b).
The matrix is sparse as indicated by the white boxes (zero-value matrix elements).
Here $N=6$ and each slice $n=1,\cdots,6$ has $M_n=4$ sites.
Note that the matrix elements due to the terminal $3$ spoil the tridiagonal block structure of $\mathbf H+\mathbf \Sigma$:
Non-zero self-energy matrix elements appear in blocks other than the first and last slices ($1$ and $6$) connecting 
simultaneously the slices $3,4$, and $5$.
In more realistic cases of wider leads, the number of non-zero self-energy matrix elements increases and they appear further 
away from the tridiagonal block structure.

The RGF method has been modified over the years to account for multiple terminals.
As mentioned in the introduction, there are some well-established schemes for multi-probe RGF in use, 
such as the cross strip \cite{Baranger1988,Baranger1990,Baranger1991} and the circular \cite{Thorgilsson2014} 
methods.
All of them, including our scheme, are faster then the full inversion.
Nevertheless, their efficiency depends strongly on the system symmetry.
The partition scheme we present here finds an optimal set of slices with minimal $M_n$ for arbitrary system geometries and
it is of very simple implementation.

%----------------------------------------------------------- F I G U R E 2
%----------------------------------------------------------- F I G U R E  2

Our MPRGF implementation relies on using the power and simplicity of the standard two-probe RGF
equations \cite{Lewenkopf2013} that is achieved by introducing an adaptive slicing scheme and a (single)
virtual lead \cite{Wimmer2009}.
This is done in two main steps.

\begin{figure}[tbp]
	\centering
		\includegraphics[width=0.83\columnwidth]{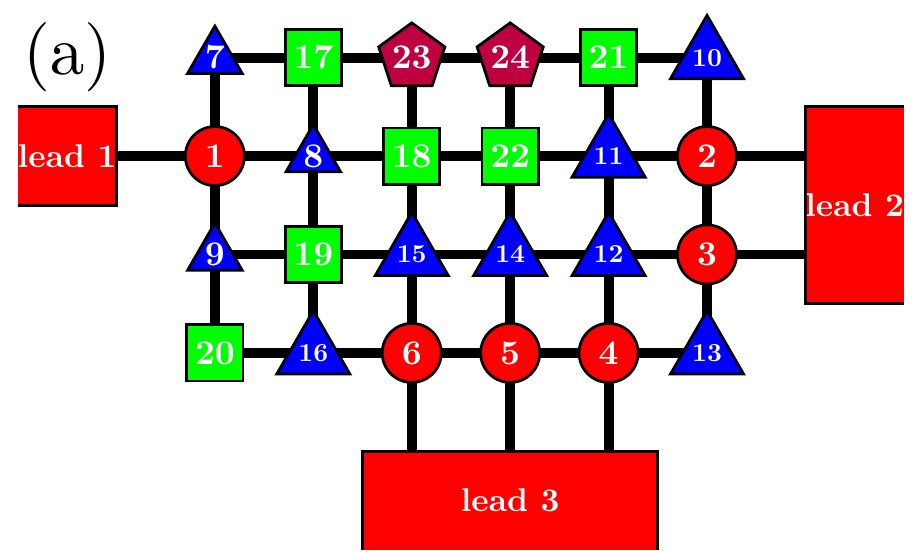}
		\includegraphics[width=0.83\columnwidth]{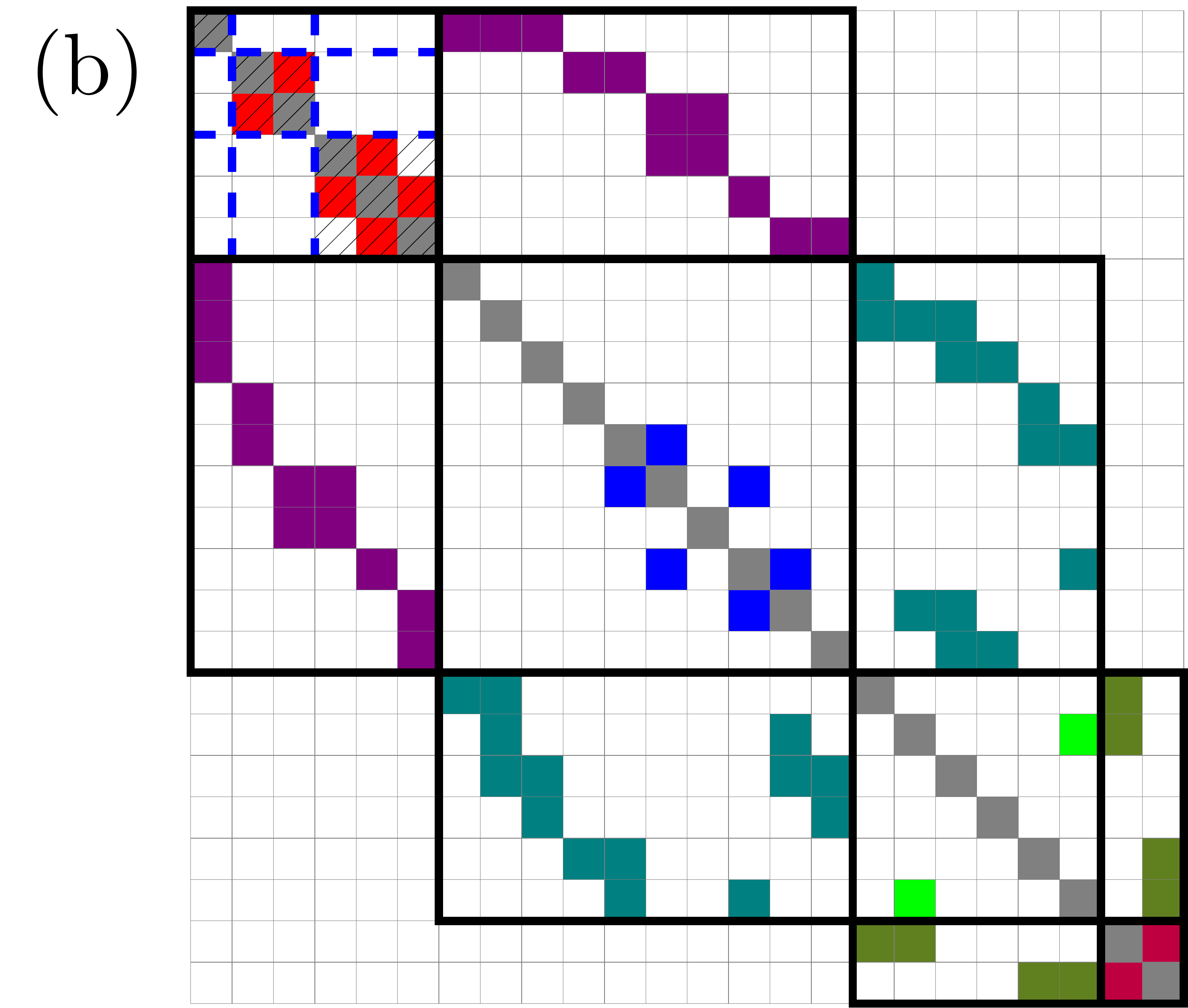}
	\caption{
	MPRGF slicing scheme applied to the system in Fig.~\ref{fig:slicing_scheme_two_probe}.
	Panel (a) shows the sites with their new labels.
	Red circles, blue triangles, green squares and purple pentagons represent sites belonging to the first, second, third and fourth slices, respectively.
	Panel (b) is the representations of the matrix $\mathbf H +\mathbf \Sigma$ corresponding to the slicing scheme in (a).
	The thick solid lines highlight the intra- and inter-slice blocks while the thick dashed lines correspond to the sub-block division of the slice $1$ (see the main text for details).
	The hatched pattern represent the contribution from the self-energy $\mathbf \Sigma$ that acts only on the diagonal sub-blocks of slice $1$ due to its ordering by leads.
	}
	\label{fig:sliced_system}
\end{figure}

(i) {\it Adaptive partition:} We start the recursion with a virtual ``left'' lead composed by all the contact sites in the leads,
which we call slice $n=0$.
We define the slice $n=1$ by the sites that are connected to any lead $\alpha=1,\cdots,N_L$, where $N_L$ is the number
of leads attached to the system.
The next slices $n=2,3,...$  are composed by the sites that are connected to sites that belong to the $n-1$ slice.
This procedure is repeated $N$ times until all lattice sites are assigned to a slice.
This scheme gives a block tridiagonal $\mathbf H$ (see mapping below) in a $N\times N$ block matrix representation.
Figure~\ref{fig:sliced_system}(a) shows the proposed slicing scheme applied to the system of Fig.~\ref{fig:slicing_scheme_two_probe}.
We use different shapes and colors to indicate the slice each site belongs.

%----------------------------------------------------------- F I G U R E 3
%----------------------------------------------------------- F I G U R E 3

(ii) {\it Site labels reassignment:}
We renumber the sites in the system according to the lead they are attached and the slice they belong in increasing order as follows.
The sites in $n=1$ are numbered in increasing order according to the leads to which they are connected to.
The number of sites connected to the lead $\alpha$ is $M_{1\alpha}$.
Thus, the total number of sites in the $n=1$ slice is $M_1=\sum_{\alpha=1}^{N_L} M_{1\alpha}$.
We divide the slice $n=1$ into $\alpha = 1, \cdots, N_L$ sub-blocks, where the $\alpha$-block contains the $M_{1\alpha}$ sites connected
to the lead $\alpha$ and has dimension $M_{1\alpha} \times M_{1\alpha}$.
The self-energy matrix $\mathbf \Sigma$ has nonzero elements only
in the sub-blocks $\mathbf \Sigma_{1,1}^{\alpha,\alpha}$ due to each real lead $\alpha$.
The numbering of sites in the slices $n\ge 2$ can follow any specific order as long as each site in slice $n$ has a higher number than any site in slice $n-1$.

For clarity, let us explicitly implement this scheme for the model system of Fig.~\ref{fig:slicing_scheme_two_probe}.
Figure~\ref{fig:sliced_system}(a) shows the result.
The sites connected to the leads that were originally numbered as $3,9,13,17,22$, and $23$ in
Fig.~\ref{fig:slicing_scheme_two_probe}(a) constitute the $n=1$, with $M_{11}=1$, $M_{12}=2$ and $M_{13}=3$ sites connected to leads $1$, $2$ and $3$, respectively, that give $M_1=6$.
Figure~\ref{fig:sliced_system}(a) indicates the sites in slice $n=1$ as red circles.
We find this site label reassignment convenient, but is certainly not unique.

Next, the sites connected to the $n=1$ are the sites with the original labels $2,4,5,7,10,14,18,19,21$ and $24$.
These $10$ sites belonging to slice $2$ are renumbered from $7$ to $16$ and shown in Fig.~\ref{fig:sliced_system}(a) as blue triangles.
Following this protocol, $n=3$ has $6$ sites ($1,6,8,11,15,20$) renumbered from $17$ to $22$ and represented as green squares, while $n=4$ contains $2$ sites ($12,16$) renumbered as $23$ and $24$ being represented by purple pentagons in Fig.~\ref{fig:sliced_system}(a).

Figure~\ref{fig:sliced_system}(b) shows the corresponding matrix structure of $\mathbf H + \mathbf \Sigma$.
Obviously the matrix has the same sparsity as before, but the size of the blocks can become larger than those 
expected in the standard RGF depending on the system.
Each diagonal sub-block of the block $n=1$ is filled by the self-energy of one lead.
% The off-diagonal sub-blocks can be populated if there are non-vanishing hopping matrix elements between sites 
% connected to different leads, which is not the case of the toy model used as an example.

In this example we see that this slicing scheme is simple and fast to implement.
It is possible to introduce optimizations to the slicing scheme, such as the one developed for two-probes in 
Ref.~\cite{Wimmer2009} based on the theory of graphs, at the cost of increasing the coding complexity.
This discussion is beyond the scope of the present work.

Now we have all the ingredients to calculate the Green's functions using the RGF method.
As standard, the free Green's functions are defined by setting the inter-slice hopping matrices $\mathbf U_{n,n+1}=0$.
By turning on the inter-slice matrices $\mathbf U_{n,n+1}$ we write a Dyson equation for the fully connected system.
We perform left and right recursions using the equations \cite{Lewenkopf2013}
\begin{align}
    \mathbf G^L_{n,n} &= \left( \mathbf E-\mathbf H_{n} - \mathbf U_{n,n-1} \mathbf G^L_{n-1,n-1} \mathbf U_{n-1,n}\right)^{-1}, \label{leftsweep}\\
    \mathbf G^R_{n,n}&=\left( \mathbf E-\mathbf H_{n} - \mathbf U_{n,n+1}\mathbf G^R_{n+1,n+1}\mathbf U_{n+1,n}\right)^{-1}, \label{rightsweep}
\end{align}
where $\mathbf G^{L}_{n,n}$($\mathbf G^{R}_{n,n}$) is the Green's function of the slice $n=1,\cdots,N$ when all the $k$ slices at its ``left'' with $k<n$ (``right'' with $k>n$) are already connected.
The recursions in Eqs.~(\ref{leftsweep}) and (\ref{rightsweep}) start at $n=1$ and $n=N$, respectively, and depend on the surface Green's functions of the virtual leads $\mathbf G^{L}_{0,0}$ and $\mathbf G^{R}_{N+1,N+1}$.
The latter are obtained by standard procedures \cite{Lewenkopf2013,LopezSancho1985}.
Since in our scheme all terminals are coupled to a single left virtual lead and the right virtual lead is uncoupled, we write
\begin{align}
    \mathbf G^L_{1,1} &= \left( \mathbf E-\mathbf H_1 -
		\mathbf \Sigma_{1,1} \right)^{-1}, \label{leftsweep1}\\
    \mathbf G^R_{N,N} &=\left( \mathbf E-\mathbf H_N \right)^{-1}, \label{rightsweepn}
\end{align}
where $\mathbf \Sigma_{1,1} \equiv \mathbf U_{1,0} \mathbf G^L_{0,0} \mathbf U_{0,1}$ is block diagonal because the real leads are decoupled, as we show in Fig.~\ref{fig:sliced_system}(b).

\begin{figure}[tbp]
	\centering
		\includegraphics[width=0.7\columnwidth]{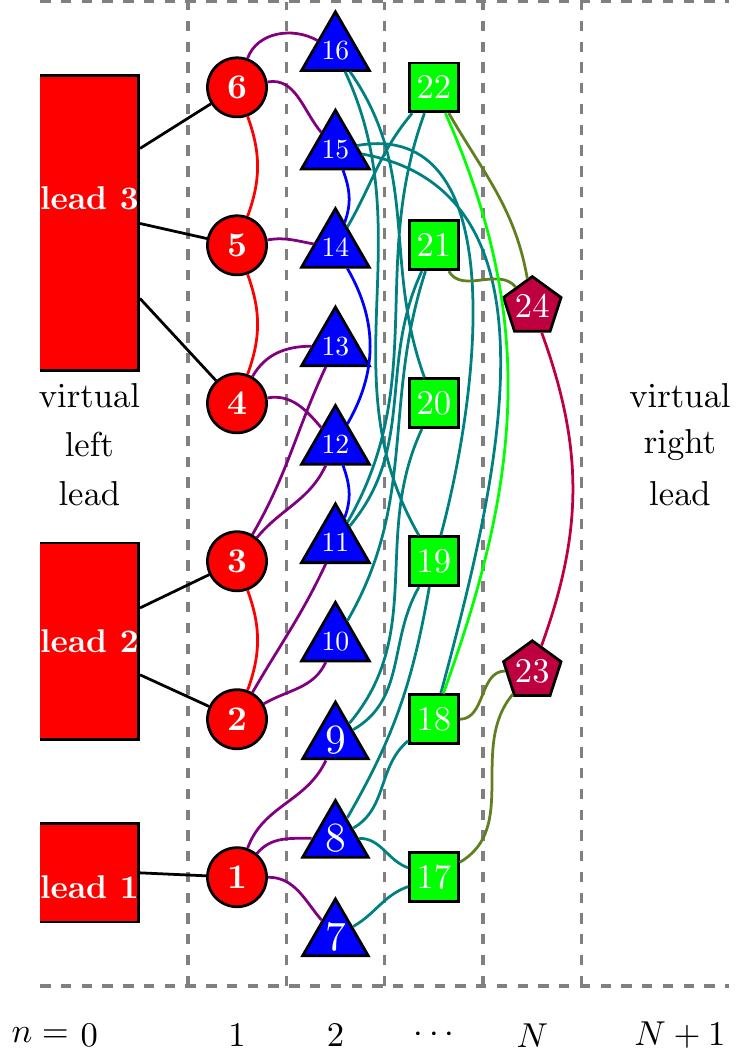}
	\caption{
		Slicing scheme for the three-terminal model system.
		The site representation (colors and shapes) is the same as in Fig.~\ref{fig:sliced_system}(a).
		The dashed lines mark the slice divisions and the solid lines represent the bonds between sites.
	The left virtual leads contains all the real leads attached to the system and the right virtual lead is empty.
	}
	\label{fig:slicing_scheme}
\end{figure}

Figure~\ref{fig:slicing_scheme} shows how the adaptive slicing scheme maps the lattice of Fig.~\ref{fig:sliced_system}(a) into an equivalent two-terminal system lattice with a virtual left lead containing all the real leads and an uncoupled virtual right lead.

The local Green's functions of the fully connected system $\mathbf G_{n,n}$ are given by \cite{Lewenkopf2013}
\begin{align}
    \mathbf G_{n,n} = \bigg( \mathbf E-\mathbf H_n
		&- \mathbf U_{n,n-1} \mathbf G^L_{n-1,n-1} \mathbf U_{n-1,n} \nonumber\\
		&- \mathbf U_{n,n+1} \mathbf G^R_{n+1,n+1} \mathbf U_{n+1,n}
		\bigg)^{-1}.
		\label{fullsweep}
\end{align}
Using Eq.~(\ref{fullsweep}) we can directly calculate local properties such as the LDOS for all the sites in the system by simply extracting the diagonal elements of $\mathbf G_{n,n}$ for all $n$ and using Eq.~(\ref{eq:ldos}).

%----------------------------------------------------------- F I G U R E 4
%----------------------------------------------------------- F I G U R E 4

To calculate the transmission matrix elements given by Eq.~(\ref{eq:transmission}) we need the Green's functions components $\mathbf G_{1,1}$ connecting sites attached to different real leads.
The Green's function $\mathbf G_{1,1}$ has dimension $M_1$ and reads
\begin{align}
	\mathbf G_{1,1} = \left(
	\begin{array}{cccc}
		\mathbf G^{1,1}_{1,1}		&\mathbf G^{1,2}_{1,1}		&\cdots	&\mathbf G^{1,N_L}_{1,1} \\
		\mathbf G^{2,1}_{1,1}		&\mathbf G^{2,2}_{1,1}		&\cdots	&\mathbf G^{2,N_L}_{1,1} \\
		\vdots						&\vdots							&  			&\vdots \\
		\mathbf G^{N_L,1}_{1,1}	&\mathbf G^{N_L,2}_{1,1}	&\cdots	&\mathbf G^{N_L,N_L}_{1,1}
	\end{array}\right)
	\label{G11subblock}
\end{align}
where each sub-block $\mathbf G^{\alpha,\beta}_{1,1}$ of dimension $M_{1\alpha}\times M_{1\beta}$ represents the 
propagator between all the $M_{1\alpha}$ sites connected to the lead $\alpha$ and all the $M_{1\beta}$ sites connected 
to the lead $\beta$.
{
Note that the sub-block division of the slice $n=1$ for $\mathbf H+\mathbf \Sigma$ naturally renders to the sub-block 
division of $\mathbf G_{1,1}$ in Eq.~(\ref{G11subblock}), see the example in Fig.~\ref{fig:sliced_system}(b).
}

If we are interested only in the total transmissions, we need to perform only the right sweep in Eq.~(\ref{rightsweep}) for $n=N,\cdots,2$ and calculate  $\mathbf G_{1,1}$ using Eq.~(\ref{fullsweep}) for $n=1$.
The calculation of other local properties such as the local transmissions $T_{jj'}^{\alpha}$ in Eq.~(\ref{eq:local_transmission}) requires the Green's functions components that connect the sites of interest $j$ and $j'$, that belong to slices $n$ and $n'$, respectively, and the sites attached to any lead $\alpha$, that belong to $n=1$.
Thus, we need the full Green's function blocks $\mathbf G_{n,1}$.

We calculate $\mathbf G_{n,1}$ by means of the extra recursions \cite{Lewenkopf2013}
\begin{align}
	\mathbf G^L_{n,1} &= \mathbf G^L_{n,n} \mathbf U_{n,n-1} \mathbf G^L_{n-1,1},\\
	\mathbf G_{n,1}   &= \mathbf G_{n,n} 	\mathbf U_{n,n-1} \mathbf G^L_{n-1,1},
\end{align}
where, as before, $n=2,\cdots,N$ and the label $L$  indicates that the Green's function $\mathbf G^L_{n,1}$ is the propagator between slices $n$ and $1$ when all the slices between them are connected.
Note that in distinction to the two-terminal RGF, here it is not necessary to compute
$\mathbf G_{N,n}$, $\mathbf G_{n,N}$ and $\mathbf G_{1,n}$.
Those matrices are not necessary because all the leads are connected only to the slice $1$ as we show in Fig.~\ref{fig:slicing_scheme}.

Once again we use the sub-block representation to write
\begin{align}
	\mathbf G_{n,1} = \left(
	\begin{array}{c}
		\mathbf G^{1  }_{n,1} 	\\
		\mathbf G^{2  }_{n,1} 	\\
		\vdots									\\
		\mathbf G^{N_L}_{n,1}
	\end{array}\right),
\end{align}
where each sub-block $\mathbf G^{\alpha}_{n,1}$ is the Green's function that connects all the $M_{1\alpha}$
sites contained in slice $n=1$ that are attached to the lead $\alpha$ to all the $M_n$ sites in the slice $n$.
For instance, by inspecting Fig.~\ref{fig:slicing_scheme} one easily concludes that $\mathbf G^{3}_{2,1}$ is a
$10 \times 3$ matrix connecting the sites $7,\cdots,16$ at slice $2$ to the sites $4,5,6$ in slice $1$ that are
attached to lead $3$.

We stress that, for simplicity, we have only discussed lattice Hamiltonians with nearest-neighbor coupling terms.
The method and equations presented here apply to any number of next-nearest-neighbors, namely, 2nd, 3rd, and so on,
which is of particular interest for tight-binding models based on maximally localized Wannier functions or related developments 
(see, for instance, Ref.~\onlinecite{Nardelli2016}.)
Obviously, the inclusion of next-nearest-neighbors increases $M_n$, since each slice $n$ is composed by all the
sites connected to the $n-1$ partition, and decreases with the number of slices $N$.
As a consequence, both the computational time and the memory usage increase with the reach of hopping integrals.

%%%%%%%%%%%%%%%%%%%%%%%%%%%%%%%%%%%%%%%%%%%%%%%%%%%%
\section{Application}
\label{sec:application}
%%%%%%%%%%%%%%%%%%%%%%%%%%%%%%%%%%%%%%%%%%%%%%%%%%%%

%\CAIO{{\bf To be done: Give an interpretation for the peak heights in $R_{xx}$ of Fig. 5.}\\}

\begin{figure}[tbp]
	\centering
		\includegraphics[width=1.00\columnwidth]{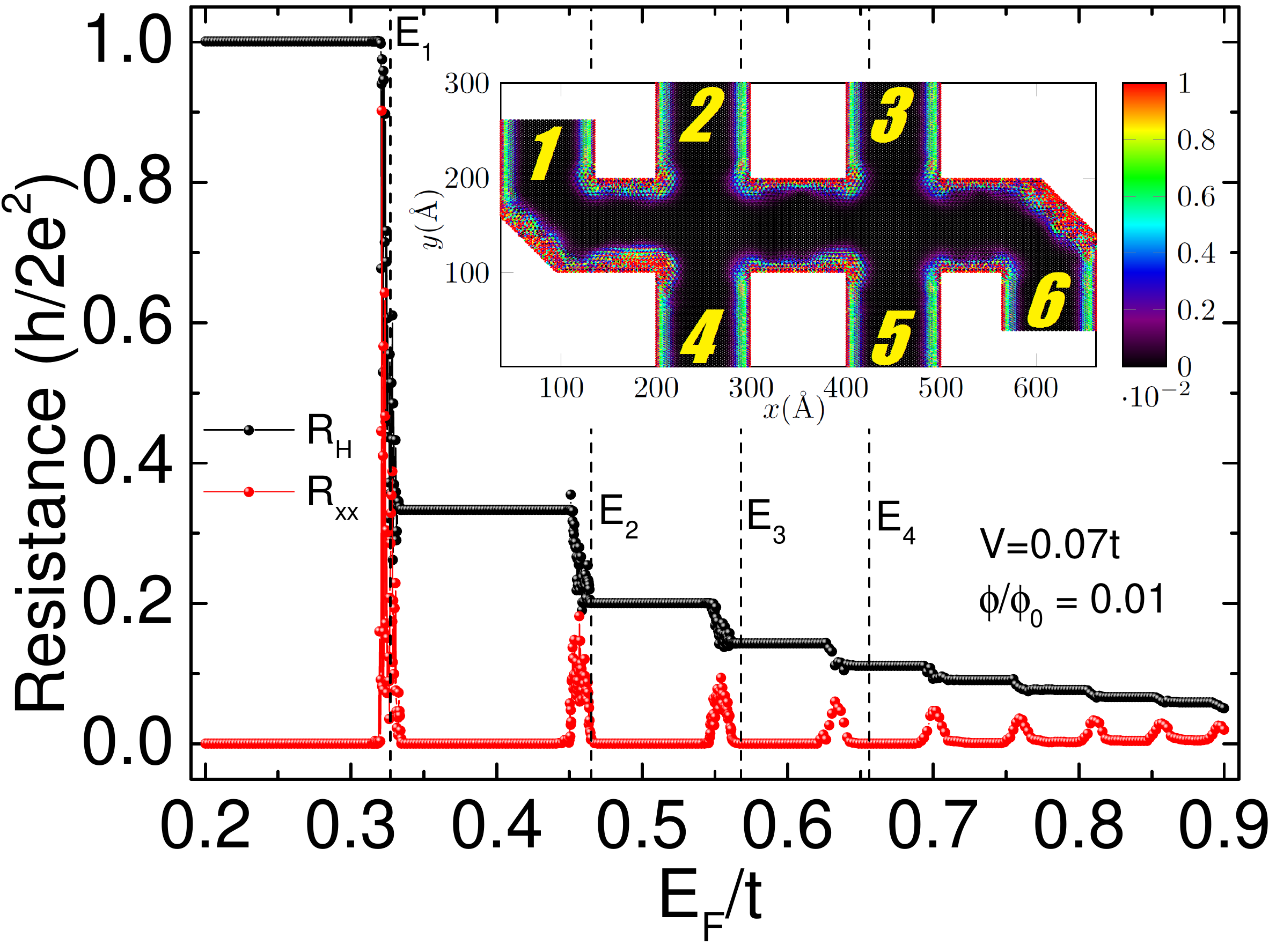}
	\caption{Longitudinal $R_{xx}$ and Hall  $R_H$ resistances as functions of the electronic energy $E_F$ for a constant magnetic flux $\phi/\phi_0=0.01$ and $V=0.07t$. The dashed lines mark the analytical value of the Landau levels $E_n$ for $n=1,2,3$ and $4$ (see text for details). The inset shows the geometry used in the MPRGF calculations. Each arm and the main branch of the Hall bar are $100$\AA\ wide. The color map shows the LDOS in arbitrary units for $E_F=0.2t$ where $1$ transport channel is open.
	}
	\label{fig:Hall_resistances_PHI001_W007}
\end{figure}

Let us illustrate the power of the method by calculating longitudinal and Hall resistances for a disordered
graphene monolayer sample submitted to a strong perpendicular magnetic field $B$ in a Hall bar geometry
with $6$ terminals, see Fig.~\ref{fig:Hall_resistances_PHI001_W007}.

The electronic properties of the system are modeled by a nearest neighbor tight-binding Hamiltonian
with disordered onsite energy \cite{CastroNeto2009,Mucciolo2010}, namely
\begin{align}
	H = -\sum_{\left\langle i,j\right\rangle} t_{i,j} c_i^\dagger c_j + \sum_j \epsilon_j c_j^\dagger c_j,
\label{H_QH}
\end{align}
where the operator $c_j^\dagger$ ($c_j$) creates (annihilates) one electron at the $p_z$ orbital of the $j$-th atom
of the graphene honeycomb lattice.
The first sum runs through nearest neighbors atoms.
The hopping matrix element between sites $i$ and $j$ is $t_{i,j} = t e^{i\varphi_{ij}}$ where $t=2.7$ eV is the
hopping integral for graphene \cite{CastroNeto2009} and $\varphi_{ij}$ is the standard Peierls phase acquired
in the path from $i$ to $j$ due to the presence of magnetic field.
The magnetic field is accounted for by a vector potential in the Landau gauge, namely, $\mathbf A = Bx\mathbf{\hat y}$.
The corresponding Peierls phase reads \cite{Lewenkopf2013}
\begin{align}
\varphi_{ij} = \frac{e}{\hbar}\int_{\mathbf r_i}^{\mathbf r_j} {\mathbf A} \cdot d\mathbf l = 2\pi\frac{\phi}{\phi_0}\frac{(x_j+x_i)(y_j-y_i)}{a_0^2 \sqrt{3}},
\end{align}
where $\mathbf r_j = x_j\hat{\mathbf x} + y_j\mathbf{\hat y}$ is the site $j$ position, $a_0 = 2.46$ \AA\ is the graphene lattice parameter, $\phi_0=h/e$ is the magnetic flux quantum and $\phi$ is the magnetic flux through one hexagon of the graphene lattice, namely $\phi=BA_H$, where $A_H=a_0^2\sqrt{3}/2$.
We use the Anderson model for the onsite disorder, where $\epsilon_j$ is randomly chosen from a uniform distribution in the interval $[-V,V]$.
The disorder strength is taken as $V=0.07t$ and the magnetic flux is $\phi/\phi_0 = 0.01$. 
The results presented below correspond to a single disorder realization.

Using the MPRGF technique described in Sec.~\ref{sec:method}
we calculate the zero-temperature conductance matrix $\mathcal G_{\alpha\beta} = (2e^2/h) T_{\alpha\beta}$ given
by Eq.~(\ref{conductance}).
We avoid spurious mode mismatch at the lead-sample interface, without the need of changing the gauge \cite{Shevtsov2012},
by using vertical leads in all six terminals of the Hall bar, see inset of Fig.~\ref{fig:Hall_resistances_PHI001_W007}.

In linear response, the multiterminal Landauer-B\"uttiker formula \cite{Buttiker1985,Buttiker1986,Ihn2010}, Eq.~(\ref{current}),
gives the current $I_\alpha$ at terminal $\alpha$ as a function of the voltages $V_\beta$ at all terminals $\beta=1,\cdots,6$.
We set
terminals $\alpha=2$ through $5$ as voltage probes with $I_2=I_3=I_4=I_5=0$ to compute the current between the terminals
$1$ and $6$, namely, $I \equiv I_1=-I_6$.
See inset of Fig.~\ref{fig:Hall_resistances_PHI001_W007}.
We obtain the longitudinal and Hall resistances using $R_{xx}=|V_4-V_5|/I$ and $R_{H}=|V_3-V_5|/I$, respectively \cite{Datta1995,Buttiker1986,Ihn2010}.

Figure \ref{fig:Hall_resistances_PHI001_W007} shows the resistances $R_{xx}$ and $R_{H}$ as functions of the electronic energy $E_F$.
We have chosen $\phi/\phi_0$ such that the system is in the quantum Hall (QH) regime.
The quantized nature of the QH effect is clearly manifest for energies where $R_{xx}=0$ and $R_H=h/2e^2(2|n|+1)$, where $2|n|+1$ is the number of propagating channels (without spin) and $n$ is the Landau Level (LL) index \cite{Datta1995,Ihn2010}.
The position of the first peak in Fig.~\ref{fig:Hall_resistances_PHI001_W007} matches the analytical value
$E_1=\sqrt{3/2}ta_0/\ell_B$ calculated using the Dirac Hamiltonian, that effectively describes the low energy
dynamics of electrons in graphene $|E_F| \ll t$ \cite{CastroNeto2009,Goerbig2011}.
When $E_F$ matches $E_n$, the energy of the Landau Level $n$, backscattering becomes available through the LL flat band channel yielding a peak in $R_{xx}$.
As expected, as one increases $|E_F|/t$, the Dirac Hamiltonian is no longer a good approximation and the numerically obtained
values of the LL energies increasingly deviate from the analytical prediction $E_n=E_1\sqrt{n}$ \cite{Brey2006}.

%----------------------------------------------------------- F I G U R E 5
%----------------------------------------------------------- F I G U R E 5

The MPRGF method is also employed to calculate the LDOS, Eq.~(\ref{eq:ldos}).
The system geometry, see inset of Fig.~\ref{fig:Hall_resistances_PHI001_W007}, 
has armchair and zigzag edges along the vertical and horizontal directions \cite{Brey2006}, respectively, 
and a rough tilted edge with no high symmetry crystallographic orientation near the 
terminals $1$ and $6$.
The inset shows that the LDOS is roughly constant along the zigzag edges. A similar 
behavior is not observed neither in
armchair nor in chiral edges. This indicates that in the QH regime the electron propagation along zigzag edges is more robust
against bulk and edge disorder than the propagation along edges with other crystallographic directions, reminiscent of the behavior
observed in the absence of an external magnetic field \cite{Mucciolo2009}.

\begin{figure}[htbp]
	\centering
		\includegraphics[width=1.0\columnwidth]{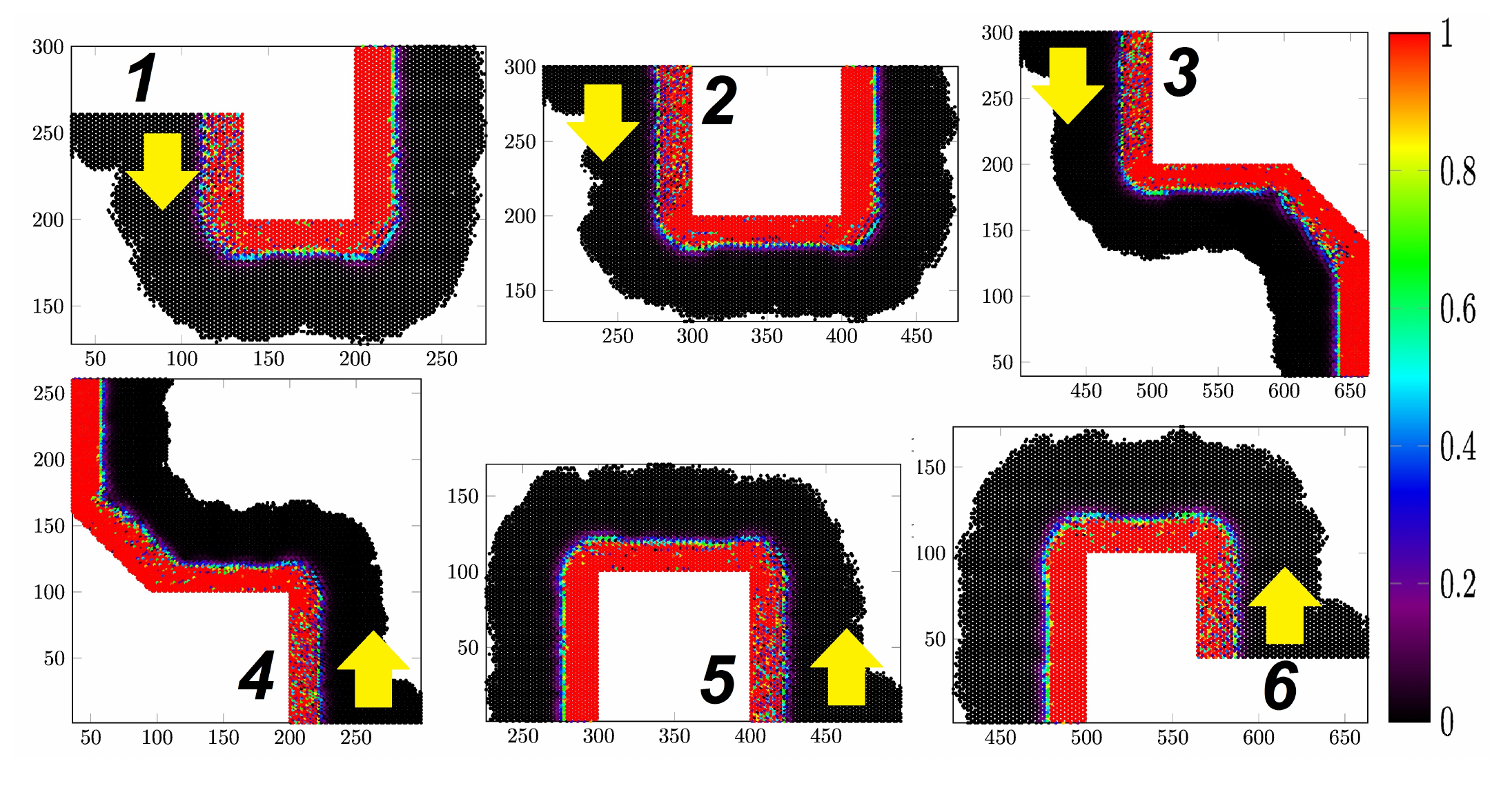}
	\caption{Local transmission in arbitrary units for $E_F=0.2t$.
	The panels consider settings where the electrons are injected from different terminals, indicated by the corresponding labels and arrows.
	We show only non vanishing transmissions.
	The parameters are the same as in Fig.~\ref{fig:Hall_resistances_PHI001_W007}.}
	\label{fig:local_transmission_QH}
\end{figure}

Figure~\ref{fig:local_transmission_QH} shows the local transmission calculated according to to Eq.~(\ref{eq:local_transmission}).
Here we set $E_F=0.2t$, corresponding to the first Hall plateau.
We find that the enhanced LDOS at opposite edges of the Hall bar observed in the inset of Fig.~\ref{fig:Hall_resistances_PHI001_W007} corresponds indeed to transmissions in opposite directions.
Electrons injected from one terminal propagate along the system edges to the next terminal on the ``left'' due to the strong magnetic field.
The edge current profile depends very weakly on which terminal the electrons are injected or on the edge crystallographic orientation, which is in contrast to the LDOS behavior in Fig.~\ref{fig:Hall_resistances_PHI001_W007}(inset).

%----------------------------------------------------------- F I G U R E 6
%----------------------------------------------------------- F I G U R E 6

%%%%%%%%%%%%%%%%%%%%%%%%%%%%%%%%%%%%%%%%%%%%%%%%%%%%
\section{Benchmark}
\label{sec:benchmark}
%%%%%%%%%%%%%%%%%%%%%%%%%%%%%%%%%%%%%%%%%%%%%%%%%%%%

\begin{figure}[tbp]
	\centering
		\includegraphics[width=1.00\columnwidth]{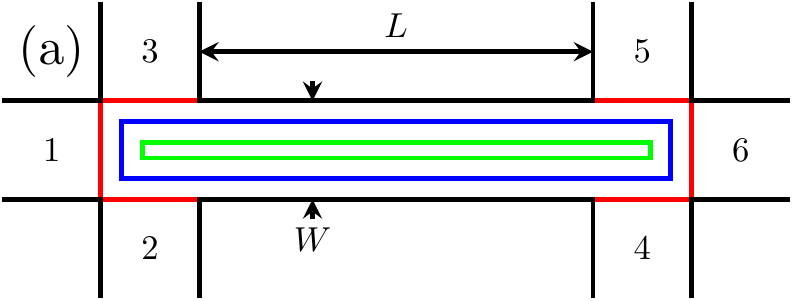}
		\includegraphics[width=1.00\columnwidth]{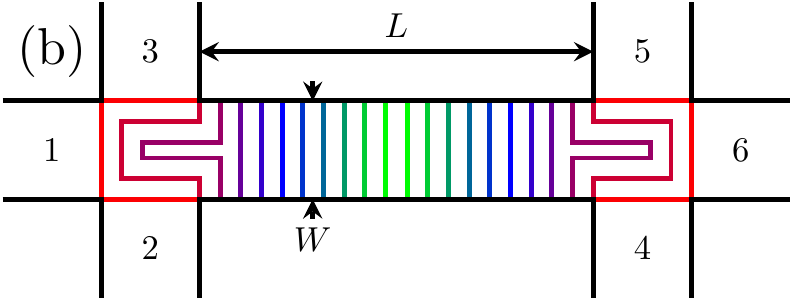}
	\caption{
	Schematic comparison between the slicing scheme used in the circular slicing and the MPRGF that we propose applied to a Hall-bar geometry. (a) The circular slicing provides slices that contain sites along both longitudinal and transverse directions, corresponding to dimensions $L$ and $W$.
	The first slice of this recursive method is the largest one, containing all the sites that are connected to the leads plus extra sites along the length $L$.
	The slices sizes decrease towards the center of the system as the last slice is the smallest one.
	(b) The MPRGF slicing scheme ensures that the first slice contains all the sites connected to the leads, rendering a smaller number of sites inside the first slice and a larger number of slices.
	}
	\label{fig:hallbars}
\end{figure}

Let us now analyze the performance and accuracy of the MPRGF method.
We compare the computational time required to calculate the transmission matrix in a six-terminals Hall bar
as depicted in Fig.~\ref{fig:hallbars} by means of direct diagonalization, circular slicing \cite{Thorgilsson2014},
and the proposed adaptive scheme.

%----------------------------------------------------------- F I G U R E  7
%----------------------------------------------------------- F I G U R E  7

The circular scheme, depicted in Fig.~\ref{fig:hallbars}(a), leads to a number $M_n$ of sites inside a slice
$n$ that depends linearly on $L$ and $W$ simultaneously.
Since the number of operations in the standard RGF scheme depends on the weight $w=\sum_{n=1}^{N} M_n^3$,
the typical runtime of a circular slicing algorithm scales as $w \propto (LW^3 + WL^3)$.
Thus, the computational time scales cubically with the largest of $L$ and $W$.
On the other hand,  in the MPRGF, the number of sites inside each cell depends mainly on $W$
while the number of slices $N$ depends mainly on $L$, which results in a computational time that scales with $LW^3$.
In both cases, the CPU time scales approximately as $L^4$ if the system has aspect ratio $W/L \approx 1$.

\begin{figure}[tbp]
	\centering
		\includegraphics[width=1.00\columnwidth]{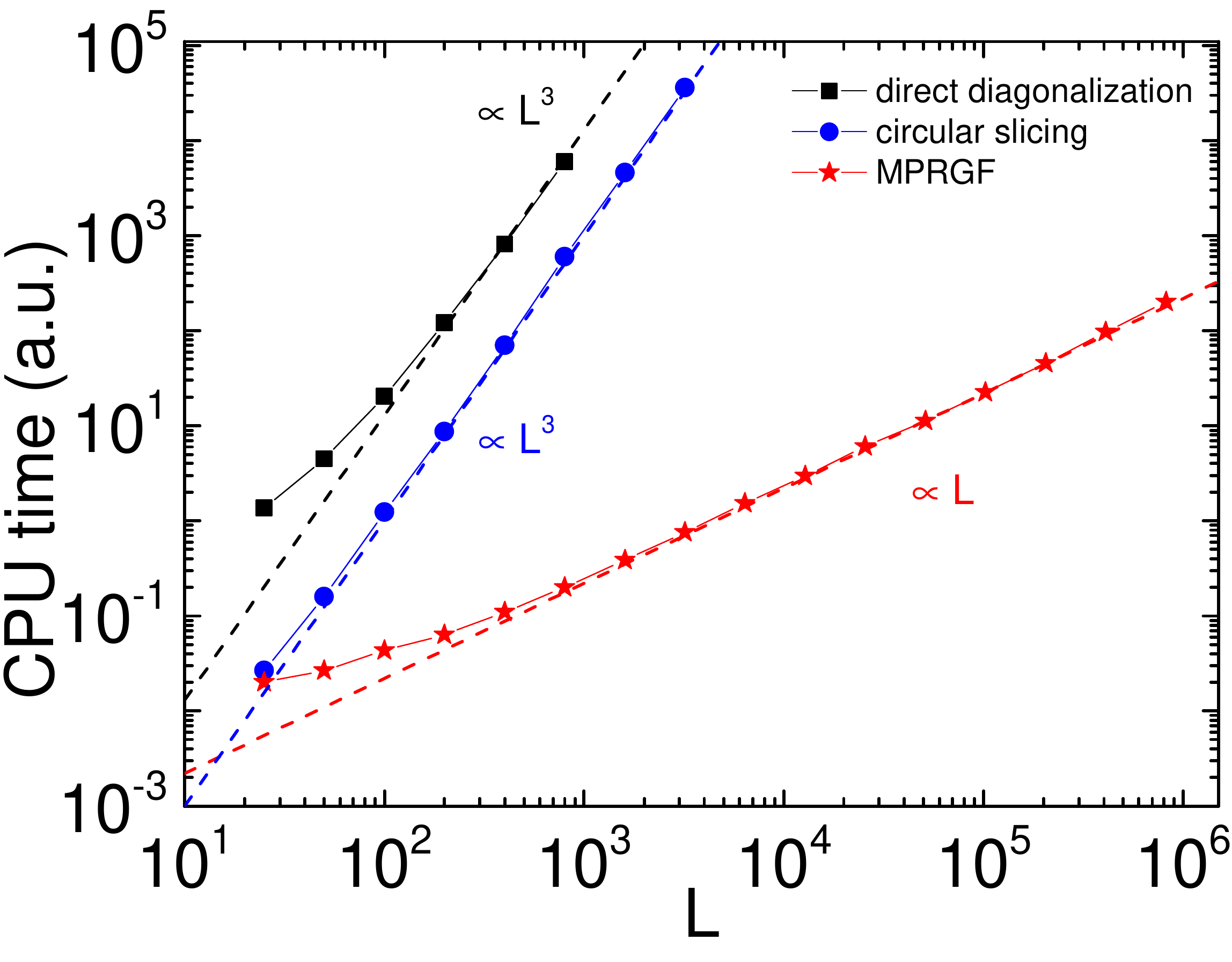}
	\caption{
	Computational time in arbitrary units (single processor) to calculate the full transmission matrix of the system depicted in Fig.~\ref{fig:hallbars} as a function of $L$ for $W=10$ using direct diagonalization (black squares), circular slicing (blue circles),
	and the proposed MPRGF (red stars).
	}
	\label{fig:cputime}
\end{figure}

Figure~\ref{fig:cputime} shows the computational time to calculate the full transmission matrix for the system in
Fig.~\ref{fig:hallbars} as a function of the length $L$, where $L\gg W$.
Here we consider a two-dimensional electron gas (2DEG) described by a discretized Hamiltonian in a square-lattice
representation with nearest-neighbors hopping matrix elements.
Using the finite differences method \cite{Datta1995,Ferry2009}, the discretization of the Schr\"odinger Hamiltonian in two-dimensions
leads to a ``hopping'' parameter $-t$, where $t=\hbar^2/(2m^*a^2)$, $m^*$ is the electron effective mass, and $a$ is the grid
spacing in both $x$ and $y$ directions. We calculate the transmission for the electron energy $E_F=0.01t$ for systems with $W=10$ sites.

We find that, for large $L$, the runtime of the adaptive scheme indeed scales linearly with $L$ while both direct
diagonalization and the circular slicing scale as $L^3$ as discussed.
The power-law dependences, which are intrinsic to the methods, render a performance to the proposed MPRGF that
is orders of magnitude better than the other codes for $L\gg W$.

Hybrid slicing schemes have been proposed to optimize the RGF method for particular system geometries. Some examples
are the cross strip \cite{Baranger1988,Baranger1990,Baranger1991} and the mixed circular \cite{Thorgilsson2014} schemes.
In these works the partition of the system region connected to the leads is designed based on the specificities of the sample
geometry, while the rest of the system is sliced by the standard method. Another nice multiprobe approach is the ``knitting" one \cite{Kazymyrenko2008},
that does not require a partition scheme. These procedures show a good computational performance,
but the coding complexity is increased. We stress that our scheme does not rely on specific features of the sample (or leads)
geometry, since it extracts from the Hamiltonian all the information needed for determining the optimal partitions.

%----------------------------------------------------------- F I G U R E  8
%----------------------------------------------------------- F I G U R E  8

\begin{figure}[htbp]
	\centering
		\includegraphics[width=1.00\columnwidth]{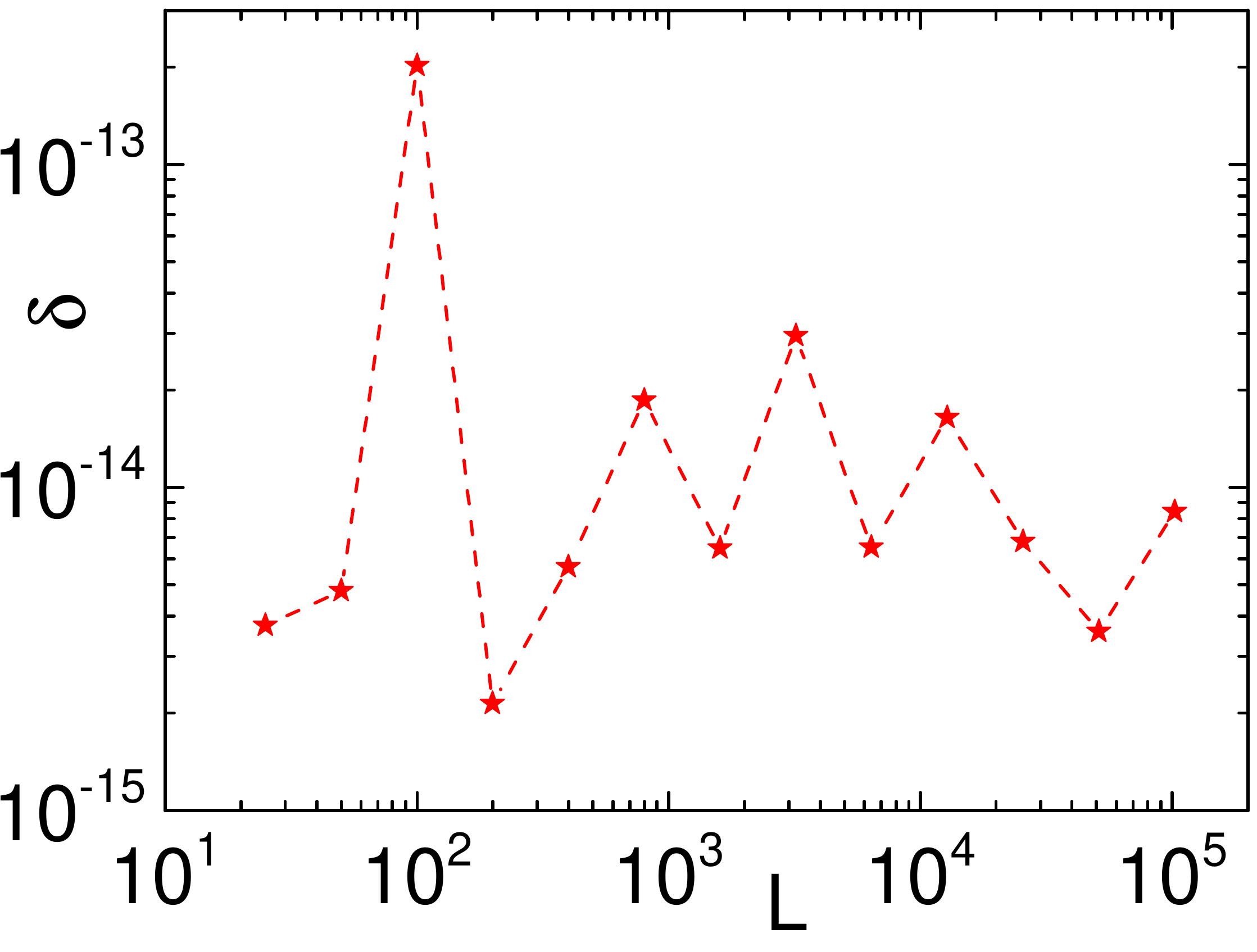}
	\caption{
	Accuracy estimate $\delta$ for the numerical calculation of the system Green's functions as a function of
	the system length $L$ for $W=10$.
	See the main text for details.
	}
	\label{fig:precision}
\end{figure}

Let us now discuss the accuracy of the adaptive scheme.
It is possible to quantify the precision of the method by comparing $(\mathbf E-\mathbf H-\mathbf \Sigma)\mathbf G$
with the unit matrix $\mathbf 1$.
This straightforward scheme cannot be used since the recursive method avoids the calculation of a large number of the
full Green's function matrix elements.
However, since
$\mathbf G_{n,n}$ and $\mathbf G_{n,1}$ are available we can
estimate the precision by evaluating
$\delta \equiv \texttt{maxval}\left[\left| \sum_n(\mathbf E-\mathbf H-\mathbf \Sigma)_{1,n}\mathbf G_{n,1} -\mathbf  1\right|\right]$,
where \texttt{maxval} returns the maximum value of the elements in the matrix.
Figure~\ref{fig:precision} shows the deviation $\delta$ as a function of the length $L$ for the system depicted in Fig.~\ref{fig:hallbars}.
By computing ${\mathbf G}$ in double precision, we find that $\delta$ does not systematically increase with $L$, supporting the
confidence on the algorithm stability, and it remains roughly within  $10^{-14} \cdots 10^{-13}$, which is 4 orders of
magnitude smaller than the deviations reported using similar methods \cite{Yang2011}.

%----------------------------------------------------------- F I G U R E  9
%----------------------------------------------------------- F I G U R E  9

We conclude this section by discussing the effect of the regularization parameter $\eta$ in the calculations.
In our approach $\eta$ is only necessary for the computation of the contact Green's function \cite{LopezSancho1985} 
and is introduced only in the first decimation loop to guarantee fast convergence.
We find that this procedure leads to a contact Green's function $\mathbf G_\alpha$ that does not depend 
on the choice of $\eta$ and minimizes the deviation between the calculated numerical value of 
$\mathbf G_\alpha$ and the one obtained by analytical expressions for 1D chains.
Since we use the ${\mathbf \Sigma}^r$ extracted from the contact Green's functions, in general it not necessary
to introduce $\eta$ for the computation of the central region Green's functions. The ``$\eta$-free" calculation 
of the system Green's function renders the numerical precision reported in this paper.

%%%%%%%%%%%%%%%%%%%%%%%%%%%%%%%%%%%%%%%%%%%%%%%%%%%%
\section{Summary}
\label{sec:discussion}
%%%%%%%%%%%%%%%%%%%%%%%%%%%%%%%%%%%%%%%%%%%%%%%%%%%%

In this paper we have put forward a multi-probe recursive Green's functions method to compute the transport properties of a
quantum phase-coherent system using the Landauer-B\"uttiker approach.

By applying the adaptive slicing scheme put forward in Sec.~\ref{sec:method}, we write the $\mathbf H+\mathbf \Sigma$ matrix
in block tridiagonal form. In this representation all leads belong to a ``left" virtual lead. Hence, the central region sites connected
to the leads belong to $n=1$ slice and there is no ``right" virtual lead attached to the slice with largest partition slice $n=N$.
This mapping allows one to use the standard RGF equations, designed to compute only the full Green's function matrix elements
necessary to calculate the transport quantities of interest, such as LDOS, local and total transmissions.

The slicing scheme we put forward allows to address multi-terminal systems with arbitrary geometries and multi-orbital
tight-binding Hamiltonians with hopping terms that include more than nearest-neighbors (at the expense of increasing CPU time).
Our method is exact, since the Green's function are calculated using the standard RGF equations, which provide a fast and robust
computational scheme that has been optimized and extensively tested over the years.
Further, it allows for the inclusion of electronic interaction via mean field approach where one needs to integrate the Green's functions in the complex plane weighted by the Fermi-Dirac distribution \cite{Lima2016, Ozaki2007, Croy2009, Areshkin2010}.

\acknowledgments
The authors acknowledge the financial support of the Brazilian funding agencies CNPq (grants: 308801/2015-6 and 400573/2017-2)
and FAPERJ (grant E-26/202.917/2015).

\bibliography{RGF}

\end{document}